\documentclass[a4paper,10pt]{article}
\usepackage[utf8]{inputenc}
\usepackage{amsmath,amssymb}
\usepackage{amsfonts}
\usepackage{epsfig}

\title{Quantum integrable  multi-well tunneling models}
\author{L H Ymai and A P Tonel  \\
Universidade Federal do Pampa, \\
Avenida Maria Anunciação Gomes de Godoy 1650\\
Bairro Malafaia, Bag\'e, RS, Brazil \\
email:leandro.ymai@unipampa.edu.br, \\
arlei.tonel@unipampa.edu.br \\
  ~~\\
A Foerster \\
Instituto de F\'isica da UFRGS, \\
Av. Bento Gon\c{c}alves 9500, \\
Agronomia, Porto Alegre, RS, Brazil
    \\
email: angela@if.ufrgs.br  \\
~~\\   
J Links\\
School of Mathematics and Physics, \\
The University of Queensland, \\ Brisbane, QLD 4072, Australia \\
email: jrl@maths.uq.edu.au}
\date{}

\begin{document}

\maketitle

\begin{abstract}
In this work we present a general construction of integrable models for boson 
tunneling in multi-well systems. 
We show how the models may be derived through the Quantum Inverse Scattering Method and solved  
by algebraic Bethe ansatz means. From the transfer matrix we 
find only two conserved operators. 
However, we construct additional conserved operators through a different method. As a consequence 
the models admit multiple pseudovacua, each associated to a set of Bethe ansatz equations. We show that all sets of Bethe ansatz equations
are needed to obtain a complete set of eigenstates.

\end{abstract}
\section{Introduction}
Since the challenging experimental realization of Bose-Einstein condensates,  our understanding about this 
state of matter has improved in both theoretical and experimental aspects. 
Nowadays this subject continues being a focus of intense investigations, with the main aim to 
understand phenomena that occur at the mesoscopic scale. It is recognized that 
exactly solvable models allow studies taking into account the whole quantum fluctuations 
that play an important role at this scale, and at ultra low temperature \cite{mtb}. 
In this direction the algebraic Bethe ansatz method has been 
an important tool to build new integrable models. Moreover this technique was already used to construct a two-mode integrable
model ({\it the two-site Bose-Hubbard model}) \cite{legg,milb,poincare} which has been used with success to describe 
experimental results \cite{will,albiez}.

Motivated by this, and the following recent developments: 
\begin{itemize}
 \item experimental efforts to investigate two-well systems with two levels in each well to study EPR entanglement \cite{he};

\item the theoretical paper {\it A bosonic multi-state two-well model} \cite{gilberto} where two solvable models in the sense 
of algebraic Bethe ansatz method are presented;

\item recent discussions about the definition of quantum integrability \cite{jonas,jean}.
\end{itemize}
Here we revisit the ref. \cite{gilberto}
to provide  a full solution for a class of multi-well tunneling models using the algebraic Bethe ansatz method. These models are defined on complete 
bipartite graphs $K_{n,m}$. 
The models are naturally associated with ($n+m$) modes, and integrability requires ($n+m$) conserved 
 operators. However the standard algebraic  Bethe Ansatz method, via the transfer matrix, provides only two of these conserved 
 operators. On the other hand, we show how to obtain the other $(n+m)-2$ additional independent conserved operators. Another important aspect is that the 
 method allows us to find a set of pseudovacua. All  pseudovacua allow 
 to build a set of Bethe states leading to a complete set of eigenvalues 
 and eigenvectors for the models.

In the next section we present the models  and 
the generalization of the algebraic Bethe ansatz technique. The approach follows the methods of ref. \cite{gilberto}, although with different notational conventions. 

\section{Integrable Hamiltonians}
We begin by introducing the Hamiltonian for ($n+m$) wells in terms of a set of canonical boson operators $a_i,\,a_i^\dagger, N_{a,i}=a_i^\dagger a_i$, $i=1,...,n$ and another set 
$b_j,\,b_j^\dagger,\,N_{b,j}=b^\dagger_j b_j$, $j=1,...,m$. The Hamiltonian reads
\begin{align}
\label{h2n}
H_{n,m} &= U(N_{A}-N_{B})^2+\mu(N_{A}-N_{B})+t\left(A^{\dagger} B+A B^\dagger\right)\nonumber \\
&=U(N_{A}-N_{B})^2+\mu(N_{A}-N_{B}) +\sum_{i=1}^n\sum_{j=1}^m t_{i,j}(a_{i}b_{j}^\dagger+a_{i}^\dagger b_{j}),
\end{align}
where we have defined $A^\dagger = \sum_{i=1}^n\alpha_i a_i^\dagger,\;B^\dagger = \sum_{j=1}^m\beta_j b_j^\dagger$, $ N_{A}= 
\sum_{i=1}^n a^\dagger_{i}a_i$ 
$N_{B}= \sum_{j=1}^m b^\dagger_{j}b_j$ and $N=N_{A}+N_{B}$. Above, the coupling  $U$ 
is the intra-well and inter-well interaction between bosons, $\mu$ is the external potential and $t_{i,j}=t\alpha_i\beta_j$ are the constant couplings for 
the tunneling amplitude. The parameters $\alpha_i,\;\beta_j$ are real numbers satisfying
\begin{align*}
 \sum_{i=1}^n \alpha_i^2=  \sum_{j=1}^m \beta_j^2=1.
\end{align*}

We will show that the above models are integrable in the sense of algebraic Bethe ansatz, but a generalization is needed. 
As we will show later, we can obtain just two conserved 
operators from the standard algebraic method through the transfer matrix, but the generalization of the method allows us to
identify $(n+m)-2$ additional constant operators.
In this sense we will show that the method presented here is a generalization for the 
algebraic Bethe ansatz method and integrability is a consequence of this generalization.

The study of the above integrable Hamiltonians is important, in part,  because it generalizes 
models that have been studied  already in the literature. To be more precise: the case $n=m=1$ is 
 the well-known {\it canonical Josephson Hamiltonian} \cite{legg,milb,poincare} which has 
 been an useful model in understanding tunneling phenomena and has been studied in many 
 aspects \cite{milb,arlei1,arlei2}. The case $n=m=2$, in the reference \cite{gilberto}, was 
 interpreted as an integrable Hamiltonian with two wells and two levels in each well. However 
 it was shown later the solution presented was not complete \cite{4well}. On the other hand non-integrable variants of these types of models were studied 
 in \cite{anglin,olsen,liberato,laha}. In the reference \cite{anglin}
 a case with two greatly different tunneling rates was studied  as 
a model for a mesoscopic quantum system in thermal contact. The 
quantum dynamics for a range of different initial conditions, 
in terms of the number of distribution among the wells and the quantum statistics, is 
presented in ref. \cite{olsen}. The tunneling 
dynamics at zero temperature was studied in ref. \cite{liberato} to investigate possible ways in which to achieve 
mass transport around a loop and persistent 
current.  In ref.\cite{laha} it was pointed out that 
an appropriate control of short-range and dipolar 
interaction may lead to novel scenarios for the dynamics of bosons in lattices, including 
the dynamical creation of mesoscopic quantum superposition, which may 
be employed in the design of Heisenberg-limited atom interferometers.

Physically one can say that the Hamiltonians (\ref{h2n}) describe Josephson 
tunneling for bosonic systems in multiple ($n+m$) wells. Besides the apparent 
simplicity, the models show a rich and beautiful mathematical structure as will be seen in the next sections.

\subsection{Some particular Hamiltonians}
From the above general Hamiltonian, for particular choices of $n,m$, we obtain the following integrable models:
\subsubsection{Two wells}
\begin{equation}
H_{1,1} = U(N_{a,1}-N_{b,1})^2+\mu(N_{a,1}-N_{b,1})+t_{1,1}(a_{1}b_{1}^\dagger+a_{1}^\dagger b_{1})
\end{equation}

\subsubsection{Three wells}
\begin{align}
H_{2,1} &= U(N_{a,1}+N_{a,2}-N_{b,1})^2+\mu(N_{a,1}+N_{a,2}-N_{b,1})\nonumber \\
&\quad+t_{1,1}(a_{1}b_{1}^\dagger+a_{1}^\dagger b_{1})+t_{2,1}(a_{2}b_{1}^\dagger+a_{2}^\dagger b_{1})
\end{align}

\subsubsection{Four wells}
\begin{align}
H_{2,2} &= U(N_{a,1}+N_{a,2}-N_{b,1}-N_{b,1})^2+\mu(N_{a,1}+N_{a,2}-N_{b,1}-N_{b,2})\nonumber \\
&\quad+t_{1,1}(a_{1}b_{1}^\dagger+a_{1}^\dagger b_{1})+t_{1,2}(a_{1}b_{2}^\dagger+a_{1}^\dagger b_{2})\nonumber \\
&\quad+ t_{2,1}(a_{2}b_{1}^\dagger+a_{2}^\dagger b_{1})+t_{2,2}(a_{2}b_{2}^\dagger+a_{2}^\dagger b_{2})
\end{align}

\begin{align}
H_{3,1} &= U(N_{a,1}+N_{a,2}+N_{a,3}-N_{b,1})^2+\mu(N_{a,1}+N_{a,2}+N_{a,3}-N_{b,1})\nonumber \\
&\quad +t_{1,1}(a_{1}b_{1}^\dagger+a_{1}^\dagger b_{1})+t_{2,1}(a_{2}b_{1}^\dagger+a_{2}^\dagger b_{1})+t_{3,1}(a_{3}b_{1}^\dagger+a_{3}^\dagger b_{1})
\end{align}

\subsubsection{Five wells}
\begin{align}
H_{3,2} &= U(N_{a,1}+N_{a,2}+N_{a,3}-N_{b,1}-N_{b,2})^2+\mu(N_{a,1}+N_{a,2}+N_{a,3}-N_{b,1}-N_{b,2})\nonumber \\
&\quad+t_{1,1}(a_{1}b_{1}^\dagger+a_{1}^\dagger b_{1})+t_{1,2}(a_{1}b_{2}^\dagger+a_{1}^\dagger b_{2})\nonumber \\
&\quad+ t_{2,1}(a_{2}b_{1}^\dagger+a_{2}^\dagger b_{1})+t_{2,2}(a_{2}b_{2}^\dagger+a_{2}^\dagger b_{2})\nonumber \\
&\quad+ t_{3,1}(a_{3}b_{1}^\dagger+a_{3}^\dagger b_{1})+t_{3,2}(a_{3}b_{2}^\dagger+a_{3}^\dagger b_{2})
\end{align}

\begin{align}
H_{4,1} &= U(N_{a,1}+N_{a,2}+N_{a,3}+N_{a,4}-N_{b,1})^2+\mu(N_{a,1}+N_{a,2}+N_{a,3}+N_{a,4}-N_{b,1})\nonumber \\
&\quad +t_{1,1}(a_{1}b_{1}^\dagger+a_{1}^\dagger b_{1})+t_{2,1}(a_{2}b_{1}^\dagger+a_{2}^\dagger b_{1})\nonumber \\
&\quad + t_{3,1}(a_{3}b_{1}^\dagger+a_{3}^\dagger b_{1})+t_{4,1}(a_{4}b_{1}^\dagger+a_{4}^\dagger b_{1})
\end{align}

\subsubsection{Six wells}
\begin{align}
H_{3,3} &= U(N_{a,1}+N_{a,2}+N_{a,3}-N_{b,1}-N_{b,2}-N_{b,3})^2 \nonumber \\
&\quad+\mu(N_{a,1}+N_{a,2}+N_{a,3}-N_{b,1}-N_{b,2}-N_{b,3})\nonumber \\
&\quad+t_{1,1}(a_{1}b_{1}^\dagger+a_{1}^\dagger b_{1})+t_{1,2}(a_{1}b_{2}^\dagger+a_{1}^\dagger b_{2})
+ t_{1,3}(a_{1}b_{3}^\dagger+a_{1}^\dagger b_{3}) \nonumber \\
&\quad+t_{2,1}(a_{2}b_{1}^\dagger+a_{2}^\dagger b_{1})+t_{2,2}(a_{2}b_{2}^\dagger+a_{2}^\dagger b_{2})
+ t_{2,3}(a_{2}b_{3}^\dagger+a_{2}^\dagger b_{3})\nonumber \\
&\quad+t_{3,1}(a_{3}b_{1}^\dagger+a_{3}^\dagger b_{1})
+ t_{3,2}(a_{3}b_{2}^\dagger+a_{3}^\dagger b_{2})+t_{3,3}(a_{3}b_{3}^\dagger+a_{3}^\dagger b_{3})
\end{align}

\begin{align}
H_{4,2} &= U(N_{a,1}+N_{a,2}+N_{a,3}+N_{a,4}-N_{b,1}-N_{b,2})^2 \nonumber \\
&\quad+\mu(N_{a,1}+N_{a,2}+N_{a,3}+N_{a,4}-N_{b,1}-N_{b,2})\nonumber \\
&\quad+t_{1,1}(a_{1}b_{1}^\dagger+a_{1}^\dagger b_{1}) +t_{1,2}(a_{1}b_{2}^\dagger+a_{1}^\dagger b_{2})  \nonumber \\
&\quad+ t_{2,1}(a_{2}b_{1}^\dagger+a_{2}^\dagger b_{1})+t_{2,2}(a_{2}b_{2}^\dagger+a_{2}^\dagger b_{2})\nonumber \\
&\quad+ t_{3,1}(a_{3}b_{1}^\dagger+a_{3}^\dagger b_{1})+t_{3,2}(a_{3}b_{2}^\dagger+a_{3}^\dagger b_{2}) \nonumber \\
&\quad + t_{4,1}(a_{4}b_{1}^\dagger+a_{4}^\dagger b_{1})+t_{4,2}(a_{4}b_{2}^\dagger+a_{4}^\dagger b_{2}).
\end{align}

\begin{align}
H_{5,1} &= U(N_{a,1}+N_{a,2}+N_{a,3}+N_{a,4}+N_{a,5}-N_{b,1})^2 \nonumber \\
&\quad+\mu(N_{a,1}+N_{a,2}+N_{a,3}+N_{a,4}+N_{a,5}-N_{b,1})\nonumber \\
&\quad+t_{1,1}(a_{1}b_{1}^\dagger+a_{1}^\dagger b_{1})+t_{2,1}(a_{2}b_{1}^\dagger +a_{2}^\dagger b_{1})+ t_{3,1}(a_{3}b_{1}^\dagger+a_{3}^\dagger b_{1}) \nonumber\\
&\quad+t_{4,1}(a_{4}b_{1}^\dagger+a_{4}^\dagger b_{1})+ t_{5,1}(a_{5}b_{1}^\dagger+a_{5}^\dagger b_{1}).
\end{align}

\begin{figure}[ht]
\begin{center}
\vspace{-2cm}
\epsfig{file=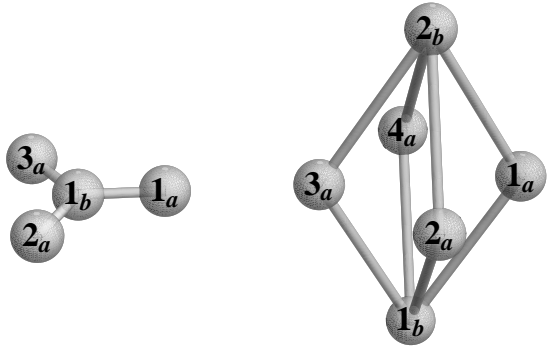,width=10.0cm,angle=0}
\vspace{-2cm}\caption{Schematic representation of the Hamiltonians $H_{3,1}$ (left) and $H_{4,2}$ (right). The spheres represent the wells, with the bonds 
indicating the tunneling between the wells. The Hamiltonian
$H_{3,1}$ is depicted with a two-dimensional geometry, while the geometry for $H_{4,2}$ is three-dimensional.}
\label{esquema}
\end{center}
\end{figure}



\section{Exact Bethe ansatz solution}
We start this section applying the Quantum Inverse Scattering Method \cite{fa1,kul,tak,fa2} to discuss the exact 
Bethe ansatz solution for the model (\ref{h2n}).

We begin with the standard $su(2)$-invariant $R$-matrix, depending on the
spectral parameter $u$:
\begin{equation}
R(u) =  \left ( \begin {array} {cccc}
1&0&0&0\\
0&b(u)&c(u)&0\\
0&c(u)&b(u)&0\\
0&0&0&1\\
\end {array} \right ),
\label{r}
\end{equation}
with $b(u)=u/(u+\eta)$ and $c(u)=\eta/(u+\eta)$.
Above, $\eta$ is a free real parameter.
It is easy to check that $R(u)$ satisfies the Yang--Baxter equation
\begin{equation}
R _{12} (u-v)  R _{13} (u)  R _{23} (v) =
R _{23} (v)  R _{13}(u)  R _{12} (u-v).
\label{ybe}
\end{equation}
Here $R_{jk}(u)$ denotes the matrix  acting non-trivially on the
$j$-th and $k$-th spaces and as the identity on the remaining space.

\subsection{General realization of Yang-Baxter algebra}
We start with the general Lax operator 
\begin{eqnarray}
L^{X}(u)=\left(\begin{array}{cc}u +\eta N_{X} & X\\X^\dagger & \eta^{-1}  \end{array}\right)\nonumber,\qquad X=A,B
\end{eqnarray}
satisfying 
\begin{equation}
R_{12}(u-v)L^X_1(u)L^X_2(v)=L^X_2(v)L^X_1(u)R_{12}(u-v),
\end{equation}
as  a result of the following algebra being satisfied
$$
[X,X^{\dagger}]= I,\;\;\;[N_X,X]=-X,\;\;\;[N_X,X^{\dagger}]=X^{\dagger}, \qquad X=A,B.
$$

Using the Lax operator presented above and the co-multiplication property \cite{jon}, we can obtain a new realization 
for the monodromy matrix that satisfies the Yang-Baxter equation through 
\begin{align*}
T(u)&= \,L^{A}(u+\omega)L^{B}(u-\omega)\\
&=\,\left( \begin{array}{cc}A(u) & B(u)\\ C(u) & D(u)
\end{array}\right).
\end{align*}
where
\begin{align}
A(u) &= (u+\omega+\eta N_{A})(u-\omega+\eta N_{B})+A B^\dagger\nonumber\\
B(u) &= (u+\omega+\eta N_{A})B+\eta^{-1} A\nonumber\\
C(u) &= (u-\omega+\eta N_{B})A^\dagger+\eta^{-1} B^\dagger\nonumber\\
D(u) &= A^\dagger B+\eta^{-2}.\nonumber
\end{align}
It can be directly shown that the monodromy matrix satisfies the Yang-Baxter 
equation
\begin{equation}
R_{12}(u-v) T_1(u) T_2(v)=
T_2(v) T_1(u)R_{12}(u-v).
\label{yba1}
\end{equation}
From this identity, there are many commutation relations between the operators $A, B, C, D$. We present those 
that are important to our discussion:
\begin{align}
\label{commutation}
A(u)C(v)&=\frac{u-v+\eta}{u-v}C(v)A(u)-\frac{\eta}{u-v}C(u)A(v)\\
D(u)C(u)&=\frac{u-v-\eta}{u-v}C(v)D(u)+\frac{\eta}{u-v}C(u)D(v).\nonumber
\end{align}

Finally, defining the transfer matrix 
\begin{equation}
\tau(u) ={\rm trace}(T(u)) = A(u)+D(u)=c_0+c_1u+c_2u^2.
\label{tm1}
\end{equation}
It follows from (\ref{yba1}) that the transfer matrix commutes for different values of the spectral parameter
$$[\tau(u),\tau(v)]=0.$$ 
Above $c_i,\;\;\;i=0,1,2$ are conserved operators given by: 
\begin{align}
 c_0&=\tau(0)=\frac{H_{n,m}}{t}+(\eta^2\frac{N^2}{4}-\omega^2+\eta^{-2})I\nonumber \\
 c_1&=\frac{d}{du}\tau(u)|_{u=0}=2\eta N\nonumber \\
 c_2&=\frac{1}{2}\frac{d^2}{du^2}\tau(u)|_{u=0}=I, \nonumber
\end{align}
where $I$ is the identity operator, and the commutation relations $[c_i,c_j]=0,\;\;i,j=0,1,2$ are satisfied.

We observe that independent of $n$ and $m$ the transfer matrix gives us just two independent conserverd operators ($H_{n,m},N)$.
 The next step is to derive the eigenvalues of the transfer matrix (\ref{tm1}). First, to apply the algebraic Bethe ansatz 
 method we have to find a pseudovacuum. In the next subsection, we show that to have a complete solution of the model the algebraic 
 method demands a set of pseudovacua.
 
 \subsection{Pseudovacua}
 
The Bethe states of the system are obtained using a set of $n$ dimensional orthonormal vectors, including  $\alpha=(\alpha_1,\alpha_2,\cdots,\alpha_n)$ and $\beta=(\beta_1,\beta_2,\cdots,\beta_m)$.
Consider $\mu_j=(\mu_{j,1},\mu_{j,2},\cdots,\mu_{j,n})$ and $\nu_j=(\nu_{j,1},\nu_{j,2},\cdots,\nu_{j,m})$ satisfying
\begin{eqnarray}
\langle \mu_j,\mu_k\rangle &=& \delta_{j,k}, \qquad  \langle \mu_j,\alpha\rangle=0, \qquad j,k = 1,2,\cdots,n-1,\nonumber \\
\langle \nu_j,\nu_k\rangle &=& \delta_{j,k}, \qquad  \langle \nu_j,\beta\rangle=0, \qquad j,k = 1,2,\cdots,m-1,\nonumber
\end{eqnarray} 
where $\langle x,y\rangle = \sum_{i=1}^{d}x_i y_i$.
Now we define the operators
\begin{align*}
\Gamma_{i}=\langle \mu_i, a \rangle, \quad  \overline{\Gamma}_{j}=\langle \nu_j, b\rangle,\quad i=1,2,\cdots,n,\qquad j=1,2,\cdots,m
\end{align*}
where $a=(a_1,a_2, \cdots, a_n)$ and $b=(b_1,b_2,\cdots,b_m)$ and the the following commutation relations are satisfied
\begin{align*}
\lbrack \Gamma_{j}^{\dagger},\overline{\Gamma}_{j}^{\dagger} \rbrack &= 0,\qquad \\
\lbrack \Gamma_{j}^{\dagger},A^\dagger \rbrack&=0, & \lbrack \overline{\Gamma}_{j}^{\dagger},B^\dagger \rbrack &=0,\\
\lbrack \Gamma_{j}^{\dagger}, A\rbrack &=0\ & \lbrack \overline{\Gamma}_{j}^{\dagger},B \rbrack &=0,\\
\lbrack\Gamma_{j},C(u)\rbrack &=0, & [\Gamma_{j},C^\dagger(u)]&=0,\\
\lbrack\overline{\Gamma}_{j},C(u)\rbrack &= \eta \overline{\Gamma}_{j} A^\dagger, & \lbrack\overline{\Gamma}_{j},C^\dagger(u)\rbrack&=\eta A\overline{\Gamma}_{j},\nonumber\\
\lbrack N_A,(\Gamma_{j}^{\dagger})^k \rbrack&= k(\Gamma_{j}^{\dagger})^k, & \lbrack N_B,(\overline{\Gamma}_{j}^{\dagger})^k \rbrack &= k(\overline{\Gamma}_{j}^{\dagger})^k.
\end{align*}
Now, denoting $\phi_{\{l;k\}}\equiv \phi_{l_1,l_2,\cdots,l_{n-1};k_1,k_2,\cdots,k_{m-1}}$, the whole set of pseudovacua can be defined as 
$$|\phi_{\{l;k\}}\rangle = \prod_{i=1}^{n-1}(\Gamma_{i}^\dagger)^{l_i}\prod_{j=1}^{m-1}(\overline{\Gamma}_{j}^\dagger)^{k_j}|0\rangle,\;\; \;\;\; r\equiv \sum_{i=1}^{n-1}l_i+\sum_{j=1}^{m-1}k_j\leq N ,$$ 
satisfying the conditions needed for the algebraic Bethe ansatz method to work, that is,
\begin{align*}
A(u)|\phi_{\{l;k\}}\rangle &= (u+\omega+\eta\sum_{i=1}^{n-1}l_i)(u-\omega+\eta\sum_{i=1}^{m-1}k_i) |\phi_{\{l;k\}}\rangle\nonumber\\
B(u)|\phi_{\{l;k\}}\rangle &= 0\nonumber\\
C(u)|\phi_{\{l;k\}}\rangle &\neq 0\nonumber\\
D(u)|\phi_{\{l;k\}}\rangle &= \eta^{-2}|\phi_{\{l;k\}}\rangle. \nonumber
\end{align*}

Denoting $\psi_{\{l;k\}}\equiv \psi_{l_1,l_2,\cdots,l_{n-1};k_1,k_2,\cdots,k_{m-1}}$, the Bethe states are given by
\begin{eqnarray}
|\psi_{\{l;k\}}\rangle=\left\{\begin{array}{ll}\prod_{i=1}^{N-r}C(v_i)\prod_{i=1}^{n-1}(\Gamma_{i}^\dagger)^{l_i}\prod_{j=1}^{m-1}((\overline{\Gamma}_{j}^\dagger)^{k_j}|0\rangle,& \mathrm{if}\,\,r<N\\
\prod_{i=1}^{n-1}(\Gamma_{i}^\dagger)^{l_i}\prod_{j=1}^{m-1}(\overline{\Gamma}_{j}^\dagger)^{k_j}|0\rangle,& \mathrm{if}\,\,r=N\end{array}\right.\nonumber
\end{eqnarray}
where $|0\rangle=|0,0,\cdots,0\rangle$ is the tensor product of the $n+m$ vacua for each mode. 

The transfer matrix eigenvalue problem is 
$$\tau(u)|\psi_{\{l;k\}}\rangle =\lambda_{\{l;k\}}(u)|\psi_{\{l;k\}}\rangle $$ where
for $r = N$ the eigenvalues are given by
\begin{equation}
 \lambda_{\{l;k\}}(u) =\left(u+\omega+\eta\sum_{i=1}^{n-1}l_i\right)\left(u-\omega+\eta\sum_{i=1}^{m-1}k_i\right)+\eta^{-2}
\end{equation}
while for $r < N$ the eigenvalues are
\begin{align*}
\lambda_{\{l;k\}}(u) &= \left(u+\omega+\eta\sum_{i=1}^{n-1}l_i\right)\left(u-\omega+\eta\sum_{i=1}^{m-1}k_i\right)\prod_{j=1}^{N-r}\frac{u-v_j+\eta}{u-v_j}\nonumber\\
&+\eta^{-2}\prod_{j=1}^{N-r}\frac{u-v_j-\eta}{u-v_j}. \nonumber
\end{align*} 
Now it is straightforward to check that the Hamiltonian (\ref{h2n}) is related to the 
transfer matrix  $\tau(u)$ (\ref{tm1}) through
\begin{equation}
H_{n,m} = t\left(\tau(u)+\omega^2-u^2-\eta^{-2}-u \tau'(0)-\frac{\tau'(0)^2}{4}\right),\nonumber
\end{equation}
where $\tau'(0) $ is the derivative in function of the spectral parameter 
and the following identification has been made for the coupling constants
$$U=-\frac{t\eta^2}{4}, \qquad \mu=-t\omega\eta.$$

The energies of the Hamiltonian (\ref{h2n}) are given by
\begin{align}
\label{h2ne}
E_{n,m}=t \left(\lambda_{\{l;k\}}(u)+\omega^2-u^2-\eta^{-2}-u \eta N-\frac{\eta^2N^2}{4}\right).
\end{align}
where $\lambda_{\{l;k\}}(u) $ is the eigenvalues of the transfer matrix 
and the set of Bethe ansatz equations (BAEs) is given by
\begin{align}
\label{bae}
\eta^2\left(v_i+\omega+\eta\sum_{i=1}^{n-1}l_i\right)\left(v_i-\omega+\eta\sum_{i=1}^{m-1}k_i\right) &= \prod_{j\neq i}^{N-r}\frac{v_i-v_j-\eta}{v_i-v_j+\eta}, \quad r<N.
\end{align} 
We remark that in the case $r = N$ there are no associated BAEs and the energy expression 
(\ref{h2ne}) takes the simple form
\begin{align}
 E_{n,m} &= t\left[\left(\sum_{i=1}^{n-1}l_i\right)\left(\sum_{i=1}^{m-1}k_i\right)\eta^2+\omega \eta\left(\sum_{i=1}^{m-1}k_i-\sum_{i=1}^{n-1}l_i\right)-\frac{\eta^2 N^2}{4}\right]\nonumber \\
 &= U\left(\sum_{i=1}^{n-1}l_i-\sum_{i=1}^{m-1}k_i \right)^2 +\mu \left(\sum_{i=1}^{n-1}l_i-\sum_{i=1}^{m-1}k_i\right).
 \label{h2nee}
\end{align}
%
%
%
It remains to show that the above method can generate a complete set of 
eigenvalues and eigenvectors for the model.

\section{Completeness and degeneracy} 
We directly diagonalize the Hamiltonian (\ref{h2n}) for two particular cases, and 
compare the results with those obtained from the algebraic Bethe ansatz. See the Appendix for details of a three-well and a six-well case, 
and also \cite{4well} for a four-well case.  
By numerical inspection we observe that for each BAE we have $N-r+1$ 
valid solutions while the other solutions  are spurious.
For fixed $r$ there are 
$$\frac{(r+n+m-3)!}{(n+m-3)!r!}$$ 
BAEs, taking into account 
the degenerate equations with $l=\sum_{i=1}^{n-1}l_i$ fixed. Considering the above comment 
that each BAE provides $N-r+1$ eigenstates of the Hamiltonian,  the total 
number of eigenstates obtained is
\begin{align}
\sum_{r=0}^N(N-r+1)\frac{(r+n+m-3)!}{(n+m-3)!r!}=\frac{(N+n+m-1)!}{(n+m-1)!N!}\nonumber
\end{align} 
which is the dimension of the Hilbert space for $N$ particles.

When $n\geq 2$ and $m\geq 2$, note that for $l=\sum_{i=1}^{n-1}l_i$ fixed it implies that $k=\sum_{i=1}^{m-1}k_i$ is also fixed. Then there 
are  
\begin{align} 
\frac{(n-2+l)!}{l!(n-2)!}\frac{(m-2+k)!}{k!(m-2)!}
\label{degen}
\end{align}
pseudovacua corresponding to the same BAE. Therefore the 
Bethe states obtained from these pseudovacua will be degenerate. This observation agrees with numerical diagonalization results given in the Appendix for the 
six-well case. When $m=1$ the number of pseudovacua with the same BAE is 
$$\frac{(n-2+l)!}{l!(n-2)!},$$
with an analogous formula for when $n=1$.

\section{Additional conserved operators}
Each Hamiltonian $H_{n.m}$ is associated with $n+m$ modes, so integrability 
requires the existence of $n+m$ independent conserved operators. The method applied above yields only two independent 
conserved operators, $H_{n,m}$ and $N$, from the transfer matrix. To obtain the other $n+m-2$ independent conserved operators,
we define the operators
\begin{align}
Q_{i}=\Gamma_{i}^\dagger\Gamma_{i}, \quad \overline{Q}_{j}=\overline{\Gamma}_{j}^\dagger\overline{\Gamma}_{j}, \quad i=1,2,\cdots,n-1\quad j=1,2,\cdots,m-1.\nonumber
\end{align}
These operators satisfy the commutation relations
\begin{align*}
 [H_{n,m},Q_{j}]=[H_{n,m},\overline{Q}_{j}]=[N,Q_{j}]=[N,\overline{Q}_{j}]&= 0 \\
 [Q_{j}, {Q}_{k}]=[Q_{j}, \overline{Q}_{k}]=[\overline{Q}_{j}, \overline{Q}_{k}]&=0,
\end{align*}
so the above $n+m-2$ operators together with the Hamiltonian $H_{n,m}$ and the number operator $N$ are the 
$n+m$ independent conserved operators for the model. 

On the other hand, the conserved operators satisfy the following commutation relations
\begin{equation}
 [Q_{j}, C(u)]=0, \;\;\;\;\ [\overline{Q}_{j}, C(u)]=0.
\end{equation}
It is seen that the Bethe states $|\psi_{\{l;k\}}\rangle$ as defined above are eigenstates of the conserved operators $Q_{j}$ and $\overline{Q}_{j}$, that is 
\begin{align}
Q_{j}|\psi_{\{l;k\}}\rangle   &= l_j|\psi_{\{l;k\}}\rangle \nonumber \\
\overline{Q}_{j}|\psi_{\{l;k\}}\rangle &=k_j|\psi_{\{l;k\}}\rangle .
\end{align}
We note that the operators $\Gamma_{j}^\dagger$ ($\Gamma_{j}$) behave 
like creation (annihillation) operators, that is, they satisfy 
the commutation relation $[\Gamma_{j},\Gamma_{j}^\dagger]=1$, while the conserved operators  
$Q_{j}=\Gamma_{j}^\dagger\Gamma_{j} $ have the action of a number operator.

\section{Conclusion}

In this work we presented a formulation for quantum integrable multi-well 
tunneling models through the Quantum Inverse Scattering Method and 
algebraic Bethe ansatz techniques.  Integrability of the Hamiltonian $H_{n,m}$ requires the existence of $n+m$ conserved 
operators, however the transfer matrix gives 
just two of them. We show how to compute the other $n+m-2$ conserved operators and associated with 
these additional conserved operators is a set of pseudovacuum states. 
Each pseudovacuum generates a a set of Bethe Ansatz equations. It was argued that all  
pseudovacua are required to obtain a complete set of eigenvalues 
and eigenvectors for each model.

As we were completing this work the preprint \cite{santos} appeared, which discusses the same class of models.  

 \vspace{1.0cm}

\centerline{{\bf Acknowledgements}}
~~\\
\noindent 
Angela Foerster and Jon Links are supported by CNPq (Conselho Nacional de Desenvolvimento Científico e Tecnológico) through Grant 450158/2016-0, and the Australian Research Council 
through Discovery Project DP150101294.~\\

\newpage


 \section*{Appendix}
 Here we compare the results of numerical diagonalization of the Hamiltonian and numerical solution of the Bethe ansatz equations for some illustrative examples. 
 \subsection*{ A.1. $H_{2,1}$ for $N=3$}
  
 In matrix form the Hamiltonian is expressible as  
  \begin{equation*}
   H_{2,1}={\scriptsize \left[ \begin {array}{cccccccccc} 9\,U+3\,\mu&t_{2,1}\,\sqrt {3
}&0&0&0&0&0&0&0&0\\ \noalign{\medskip}t_{2,1}\,\sqrt {3}&U+\mu&2
\,t_{2,1}&0&t_{1,1}&0&0&0&0&0\\ \noalign{\medskip}0&2\,t_{2,1}&U-\mu&t_{2,1}\,\sqrt {3}&0&t_{1,1}\,\sqrt {2}&0
&0&0&0\\ \noalign{\medskip}0&0&t_{2,1}\,\sqrt {3}&9\,U-3\,\mu&0&0
&t_{1,1}\,\sqrt {3}&0&0&0\\ \noalign{\medskip}0&t_{1,1}&0&0
&9\,U+3\,\mu&t_{2,1}\,\sqrt {2}&0&0&0&0\\ \noalign{\medskip}0&0&
t_{1,1}\,\sqrt {2}&0&t_{2,1}\,\sqrt {2}&U+\mu&t_{2,1}\,\sqrt {2}&t_{1,1}\,\sqrt {2}&0&0\\ \noalign{\medskip}0&0&0&t_{1,1}\,\sqrt {3}&0&t_{2,1}\,\sqrt {2}&U-\mu&0&2\,t_{1,1}&0\\ \noalign{\medskip}0&0&0&0&0&t_{1,1}\,\sqrt {2}&0&9\,
U+3\,\mu&t_{2,1}&0\\ \noalign{\medskip}0&0&0&0&0&0&2\,t_{1,1}&t_3&U+\mu&t_{1,1}\,\sqrt {3}
\\ \noalign{\medskip}0&0&0&0&0&0&0&0&t_{1,1}\,\sqrt {3}&9\,U+3\,
\mu\end {array} \right]} .
  \end{equation*}
  Choosing the coupling parameter values 
  $$U=1, \quad \mu=0.5, \quad t=-0.5, \quad \alpha_1=\alpha_2=\frac{1}{\sqrt{2}}, \quad \beta_1=1$$
  we obtain the ordered eigenspectrum below through numerical diagonalization:
\begin{equation}
\label{en3}
 \begin {array}{c}E_1= - 0.207868448700000014\\ \noalign{\medskip}
E_2 = 0.123564632300000005\\ \noalign{\medskip} E_3 =1.47230743099999994
\\ \noalign{\medskip}E_4 = 1.82091556600000004\\ \noalign{\medskip}
E_5 = 2.01646645300000005\\ \noalign{\medskip}E_6 = 7.60790053000000022
\\ \noalign{\medskip}E_7 = 10.5000000000000000\\ \noalign{\medskip}
E_8 = 10.5276925699999993\\ \noalign{\medskip} E_9 =10.5555198000000008
\\ \noalign{\medskip} E_{10} =10.5835014699999999.\end {array}
  \end{equation}
We compare these results with those obtained from the Bethe ansatz equations, which are displayed in 
Table \ref{table2}. Note that in this table 
we do not present  spurious solutions, such as those where roots of the BAE are equal. The column $5$ shows the resulting by the Bethe ansatz equations 
by comparing with the exact diagonalization (\ref{en3}). It is seen that there is a one-to-one correspondence between the results of the two approaches.

\begin{table}[h]
\centering
\caption{BAE solutions of $H_{2,1}$ for $N=3$}
\vspace{0.5cm}
\begin{tabular}{|l|l|l|l|l|}
\hline
l & pseudovacuum & BAE & BAE solution & energy \\ 
\hline                               
0 & $|0\rangle $      & $ \eta^2(v_1^2-\omega^2)=\left(\frac{v_1-v_2-\eta}{v_1-v_2+\eta}\right)\left(\frac{v_1-v_3-\eta}{v_1-v_3+\eta}\right)$ & \begin{minipage}{3cm} \scriptsize{$v_1 = 0.3643816442\\ v_2 = -0.4444604439\\ v_3 = -1.894256353 $}\end{minipage}& $E_1$\\ \cline{4-5}
  &                   & $\eta^2(v_2^2-\omega^2)=\left(\frac{v_2-v_1-\eta}{v_2-v_1+\eta}\right)\left(\frac{v_2-v_3-\eta}{v_2-v_3+\eta}\right)$ &\begin{minipage}{3cm} \scriptsize{$v_1 =0.3359698043 \\ v_2 =-0.3347796411 \\ v_3 =-3.548367609  $}\end{minipage} & $E_5$\\ \cline{4-5}
   &                   & $\eta^2(v_3^2-\omega^2)=\left(\frac{v_3-v_1-\eta}{v_3-v_1+\eta}\right)\left(\frac{v_3-v_2-\eta}{v_3-v_2+\eta}\right)$ &\begin{minipage}{3cm} \scriptsize{$v_1 =0.3535693555 \\ v_2 =-2.47636450 \\ v_3 =-5.378123245  $}\end{minipage} & $E_6$\\ \cline{4-5}
   &                  &                                                                                              &\begin{minipage}{3cm} \scriptsize{$v_1 =-0.3535609123 \\ v_2 = -3.182700505 \\ v_3 = -6.068724579  $}\end{minipage} & $E_{10}$\\ \cline{4-5}
\hline
1 & $\Gamma^\dagger|0\rangle $ & $ \eta^2(v_1+\omega+\eta)(v_1-\omega)=\frac{v_1-v_2-\eta}{v_1-v_2+\eta}$ &\begin{minipage}{3cm} \scriptsize{$v_1 =-2.206922679 \\ v_2 = 0.3517823376$}\end{minipage} & $E_{2}$ \\ \cline{4-5}
  &                            & $ \eta^2(v_2+\omega+\eta)(v_2-\omega)=\frac{v_2-v_2-\eta}{v_2-v_2+\eta}$ &\begin{minipage}{3cm} \scriptsize{$v_1 =-3.413951691 \\ v_2 =0.3586029944 $}\end{minipage}  & $E_{4}$  \\ \cline{4-5}
  &                            &                                                                          &\begin{minipage}{3cm} \scriptsize{$v_1 = -6.049425376\\ v_2 =-3.182221204$}\end{minipage}  &  $E_{9}$         \\         
  \hline
2 & $\Gamma^\dagger\Gamma^\dagger|0\rangle $  &  $  \eta^2(v_1+\omega+2\eta)(v_1-\omega)=1 $  &\scriptsize{ $v_1 = -6.010407638$} & $E_8$ \\\cline{4-5}
  &                                           &                                               &\scriptsize{ $v_1 = 0.3731349939$ } &  $E_3$  \\
\hline
3 & $\Gamma^\dagger\Gamma^\dagger\Gamma^\dagger|0\rangle $  & $\nexists $  & $\nexists $ &$E_7$ \\
\hline
\end{tabular}
\label{table2}
\end{table}  

\newpage 

\subsection*{A.2  $H_{3,3}$ for $N=3$}
We set the parameters as
\begin{align*}
 N&=3, &U&=1.3,& \mu&=0.5,& t&=-3.7,  \\
 \phi_1&=\pi/3, &\theta_1&=\pi/6, &\phi_2&=\pi/4, & \theta_2&=\pi/7, 
 \end{align*}
 with
 \begin{align*}
 \alpha_1&=\sin\phi_1\cos\theta_1, & 
\alpha_2&=\sin\phi_1\sin\theta_1, &\alpha_3&=\cos\phi_1, \\ 
\beta_1&=\sin\phi_1\cos\theta_1, &\beta_2&=\sin\phi_2\sin\theta_2, &\beta_3&=\cos\phi_2.
\end{align*}
Ignoring spurious solutions, Table \ref{table3} lists the spectrum obtained by numerically solving the Bethe ansatz equations. These are grouped for the different sectors determined 
by the various choices of pseudovacua for each fixed $l$ and $k$. The results have been compared with those obtained by direct numerical diagonalization, and it was again 
found that there is a one-to-one correspondence. In particular, the degeneracies are found to be in complete agreement with the formula (\ref{degen}).

\begin{table}
\centering
\caption{Energy spectrum obtained by the BAE for $H_{3,3}$ for $N=3$}
{
\vspace{0.5cm}
\begin{tabular}{|c|c|c|c|c|l|}
\hline
$r$ & $l_1$ & $l_2$ & $k_1$ & $k_2$  & energy\\
\hline
0&0 & 0 & 0 & 0 & \begin{minipage}{6cm}
 $ \\-8.20422, 3.59673, 13.3926, 17.2149 \\$\end{minipage}\\
\hline \hline
1&1 & 0& 0& 0& $-4.70392, 4.99326, 15.5107 $ \\
\cline{2-5}
&0 & 1& 0& 0& \\
\hline \hline
1&0 & 0 & 1 & 0 & $-5.01626, 4.81132, 13.0049 $\\
\cline{2-5}
&0 & 0 & 0 & 1 &\\
\hline \hline
2&1 & 0 & 1 & 0 & \\
\cline{2-5}
&1 & 0 & 0 & 1 & -2.43363, 5.03363\\
\cline{2-5}
&0 & 1 & 1 & 0 & \\
\cline{2-5}
&0 & 1 & 0 & 1 & \\
\hline \hline
2&1 & 1 & 0 & 0 & \\
\cline{2-5}
&2 & 0 & 0 & 0 & 0.704413, 14.2956 \\
\cline{2-5}
&0& 2 & 0 & 0 & \\
\hline \hline
2&0 & 0 & 1 & 1 & \\
\cline{2-5}
&0 & 0 & 2 & 0 & -0.481639, 11.4816 \\
\cline{2-5}
&0& 0 & 0 & 2 & \\
\hline \hline
3&1 & 1 & 1 & 0 & \\
\cline{2-5}
&1 & 1 & 0 & 1 & \\
\cline{2-5}
&2 & 0 & 1 & 0 & 1.8 \\
\cline{2-5}
&2 & 0 & 0 & 1 & \\
\cline{2-5}
&0 & 2 & 1 & 0 & \\
\cline{2-5}
&0 & 2 & 0 & 1 & \\
\hline\hline
3&1 & 0& 2 & 0 & \\
\cline{2-5}
&0 & 1 & 2 & 0 & \\
\cline{2-5}
&1 & 0 & 0 & 2 & 0.8 \\
\cline{2-5}
&0 & 1 & 0 & 2 & \\
\cline{2-5}
&1 & 0 & 1 & 1 & \\
\cline{2-5}
&0 & 1 & 1 & 1 & \\
\hline \hline
3&3 & 0& 0 & 0 & \\
\cline{2-5}
&0 & 3 & 0 & 0 & 13.2\\
\cline{2-5}
&2 & 1 & 0 & 0 & \\
\cline{2-5}
&1 & 2 & 0 & 0 & \\
\hline \hline
3&0 & 0& 3 & 0 & \\
\cline{2-5}
&0 & 0 & 0 & 3 & 10.2\\
\cline{2-5}
&0 & 0 & 1 & 2 & \\
\cline{2-5}
&0 & 0 & 2 & 1 & \\
\hline
\end{tabular}
\label{table3}
} 
\end{table}
\newpage

\end{document}